\documentclass[11pt]{article}
\usepackage{moriond,epsfig}

\def\Journal#1#2#3#4{{#1} {\bf #2}, #3 (#4)}

\def\NPB{{\em Nucl. Phys.} B}

\def\PRL{\em Phys. Rev. Lett.}
\def\PRD{{\em Phys. Rev.} D}
\def\PRC{{\em Phys. Rev.} C}

\def\be{\begin{equation}}
\def\ee{\end{equation}}
\def\bea{\begin{eqnarray}}
\def\eea{\end{eqnarray}}

\begin{document}
\vspace*{4cm}
\title{Forward physics : from SPS to LHC, what can we learn from air showers ?}

\author{ TANGUY PIEROG }

\address{Forschungszentrum Karlsruhe, Institut f\"ur Kernphysik, \\
Postfach 3640, 76021 Karlsruhe \\
Germany}

\maketitle\abstracts{
Since recent RHIC data and the development of new theories for small x physics,
a new interest appeared for forward physics. At LHC, a correct description of
multiple parton interactions will be crucial to understand all the results. On
the other hand, forward particle production and multiple interactions are the
key points of air shower development. That's why air shower measurements done
by precise experiments like KASCADE can help to
understand high energy interactions, using hadronic models which are able to
reproduce both accelerator and air shower data. In the framework of the EPOS
model, we will show what constraints can be fixed by air shower experiment on
particle production from SPS to LHC energies.}

\section{Introduction}

In this paper, we discuss how the comparison of extensive air shower (EAS) simulations based on EPOS, 
could provide new constraints for a model used in particle physics. EPOS is a hadronic 
interaction model, which does very well
compared to RHIC data~\cite{Bellwied}, and also other
particle physic experiments (especially SPS experiments at CERN). 
But used in the air shower simulation
program CORSIKA~\cite{corsika}, some results where
in contradiction with KASCADE data~\cite{KASCADE-EPOS}, while it was better for
other cosmic ray experiments~\cite{Glushkov:2007gd}.

Due to the constrains of particle physics, air shower simulations using EPOS 
present a larger number of muons at ground~\cite{Pierog:2006qv}. On the other hand,
the constrains given by cosmic ray experiments 
can compensate the lack of accelerator data in some given kinematic regions 
(very forward) and can be used to improve hadronic interaction models and in particular EPOS.

\section{EPOS Model}

One may consider the simple parton model to be the basis of high energy 
hadron-hadron interaction models, which can be seen as an exchange of a 
``parton ladder'' between the two hadrons.

In EPOS, the term ``parton ladder'' is actually meant to contain two parts \cite{nexus}:
the hard one, as discussed above, and a soft one, which is a purely
phenomenological object, parameterized in Regge pole fashion.

In additions to the parton ladder, there is another source of particle production:
the two off-shell remnants.
We showed in ref. \cite{nex-bar}
that this {}``three object picture'' can
solve the {}``multi-strange baryon problem'' of conventional
high energy models, see  ref. \cite{sbaryons}.

Hence EPOS is a consistent quantum mechanical multiple scattering approach
based on partons and strings~\cite{nexus}, where cross sections
and the particle production are calculated consistently, taking into
account energy conservation in both cases (unlike other models where
energy conservation is not considered for cross section calculations~\cite{hladik}).
Nuclear effects related to Cronin transverse
momentum broadening, parton saturation, and screening have been introduced
into EPOS~\cite{splitting}. Furthermore, high density effects leading
to collective behavior in heavy ion collisions are also taken into
account~\cite{corona}.

Energy momentum sharing and remnant treatment are the key points of the model
concerning air shower simulations because they directly influence the 
multiplicity and the inelasticity of the model. At very high energies or high 
densities, the so-called non-linear effects described in~\cite{splitting} are 
particularly important for the extrapolation for EAS and it's one of the parts 
which has been changed in EPOS~1.99 in comparison with the previous version~1.61.

\subsection{Cross section}

We learned from KASCADE data~\cite{KASCADE-EPOS}, that the energy carried by
hadrons in EPOS~1.61 simulations is too low. It means than the showers are too old
when they reach ground and it was due to a problem in the calculation of the nuclear
cross section and to a too large remnant break-up at high energy (leading to a high 
inelasticity).

\begin{figure}[hbp]
\begin{minipage}{0.55\textwidth}
\begin{center}
\includegraphics[width=0.6\textwidth]{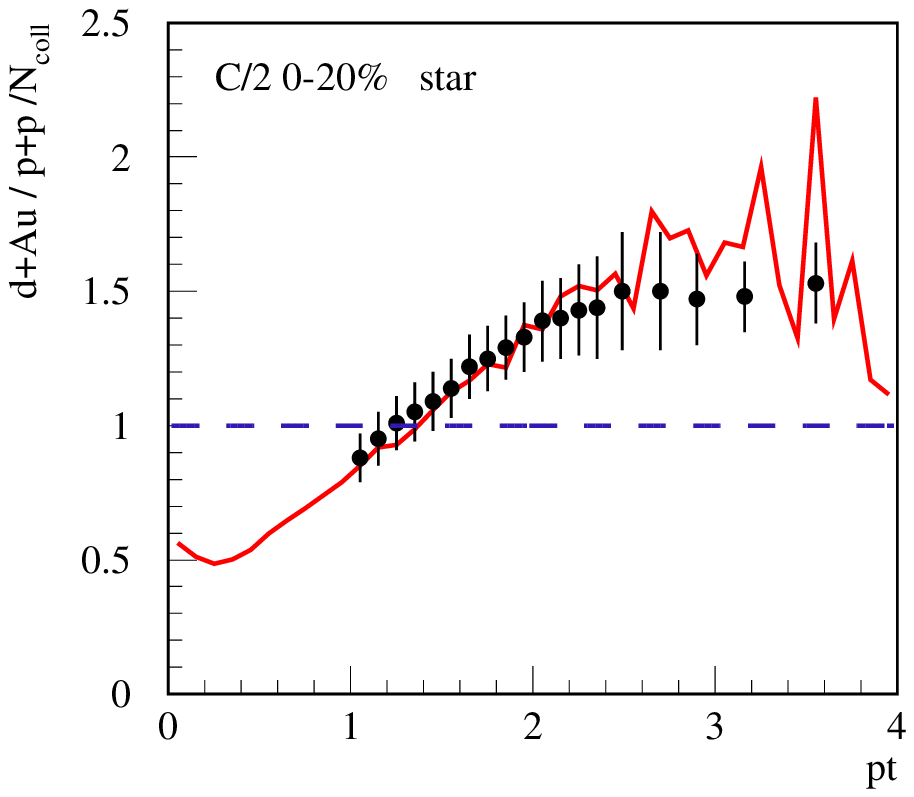}
\end{center}
\caption{Ratio of the most central deuteron-gold collisions over proton-proton normalized 
by the Glauber number of binary collisions for the
 pt distribution of charged particles at~200\,GeV for EPOS (line) and compared to data~\protect\cite{rhic} (points)\label{fig-rhic}}
\end{minipage}\hfill
\begin{minipage}{0.42\textwidth}
\begin{center}
\includegraphics[width=0.786\textwidth]{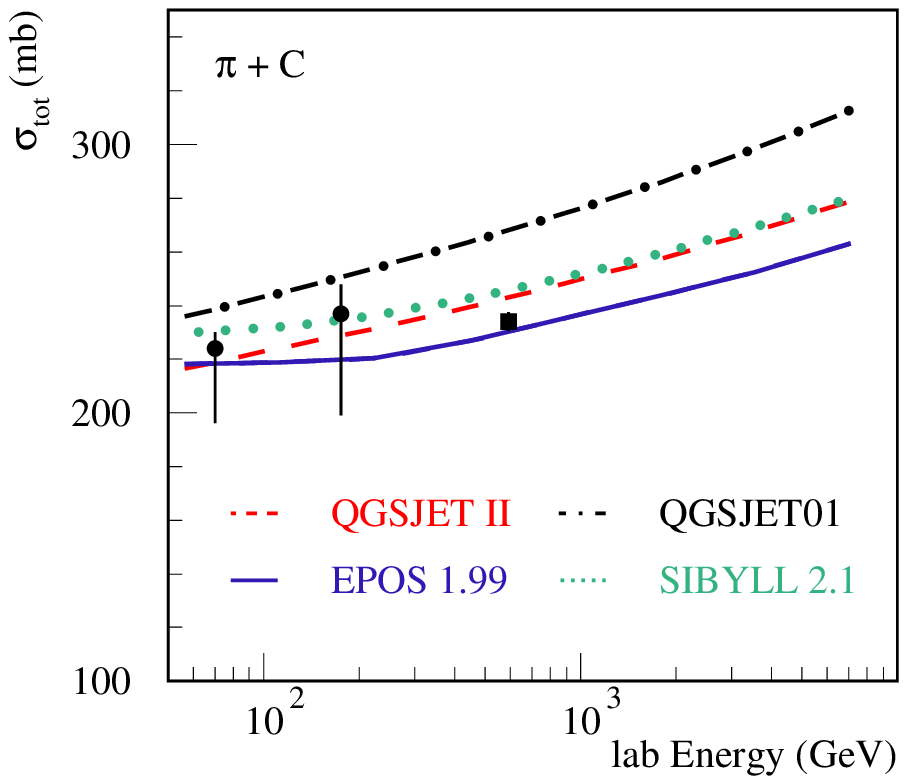}
\end{center}
\caption{Total cross section of $\pi$-carbon interactions. EPOS~1.99, QGSJET~II, QGSJET01
 and SIBYLL~2.1 hadronic interaction models (lines) are compared to data~\protect\cite{Dersch:1999zg}
 (points) \label{fig-sigC}}
\end{minipage}
\end{figure}

To improve the predictive power of the model, the effective treatment of non-linear
effects describe in \cite{splitting} has been made consistent to describe both proton-proton,
hadron-nucleus and nucleus-nucleus data with a unique saturation scale which can be fixed
thanks to proton-proton cross section and Cronin effect in dAu collisions at RHIC as 
shown fig.~\ref{fig-rhic}. Details will be published in a dedicated article.

\begin{figure}[hptb]
\begin{minipage}{0.47\textwidth}
\begin{center}
\includegraphics[width=1.\textwidth]{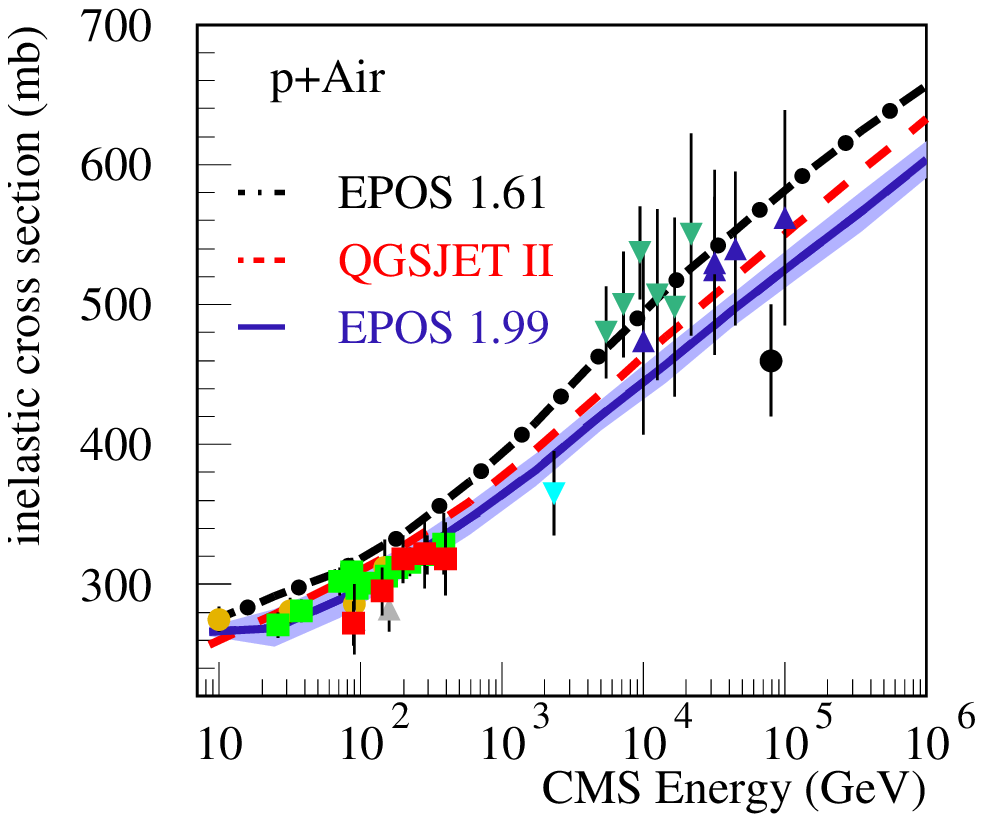}
\end{center}
\caption{Inelastic cross section of proton-air interactions. EPOS~1.99, QGSJET~II, and EPOS~1.61
 hadronic interaction models (lines) are compared to data of air shower experiment (points). \label{fig-sigAir}}
\end{minipage}
\hfill
\begin{minipage}{0.47\textwidth}
\begin{center}
\includegraphics[width=1.\textwidth]{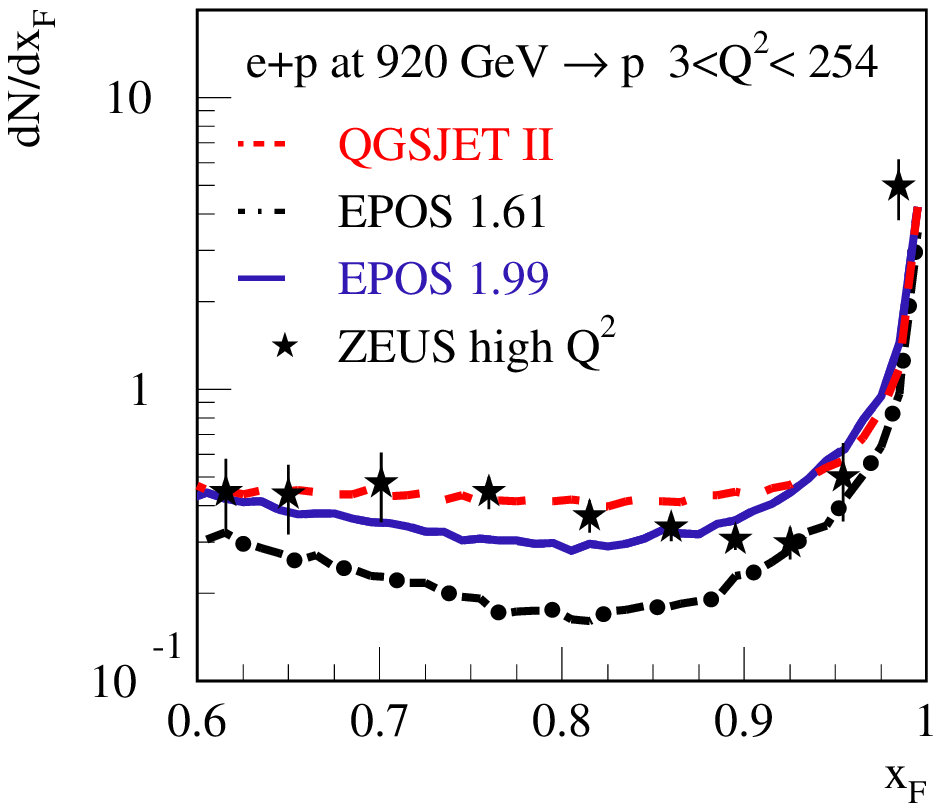}
\end{center}
\caption{Proton longitudinal momentum fraction $x_{\rm L}$ distribution in 
e-p interactions. EPOS~1.99 (full), QGSJET~II (dashed), and
EPOS~1.61 (dashed-dotted) models are compared to 
data~\protect\cite{Chekanov:2002yh} (stars). \label{fig-zeus}}
\end{minipage}
\end{figure}

The EPOS~1.99 (full line) pion-carbon total cross section is shown Fig~\ref{fig-sigC}.
It is now in very good agreement with the data~\cite{Dersch:1999zg} while 
the other hadronic interaction models used
for air shower physics QGSJET01~\cite{qgsjet01} (dashed-dotted line), QGSJET~II~\cite{qgsjetII} (dashed line) 
and SIBYLL~\cite{sibyll} (dotted line) overestimate the pion-carbon cross-section for energies above
100\,GeV. In fig~\ref{fig-sigAir}, 
the extrapolation to proton-air data up to the highest energies is shown in comparison with 
measurement from cosmic ray experiments. The surface around the line for EPOS~1.99 represents 
the uncertainty 
due to the definition of the inelastic cross section as measured by cosmic ray 
experiments. The difference between the top and the bottom of the area is the part 
of the cross-section where secondary particles are produced without changing the projectile 
(target diffraction). So any cross section chosen in this band would give the same result in term
of air shower development. Cross section of other models include this target diffraction 
(top of the band). In comparison with EPOS~1.61 (dashed-dotted line), the EPOS~1.99 cross section
 has been notably reduced.

\subsection{Inelasticity}

\begin{figure}[htbp]
\begin{minipage}{0.49\textwidth}
\vspace*{5mm}
\begin{center}\includegraphics[width=0.98\textwidth]{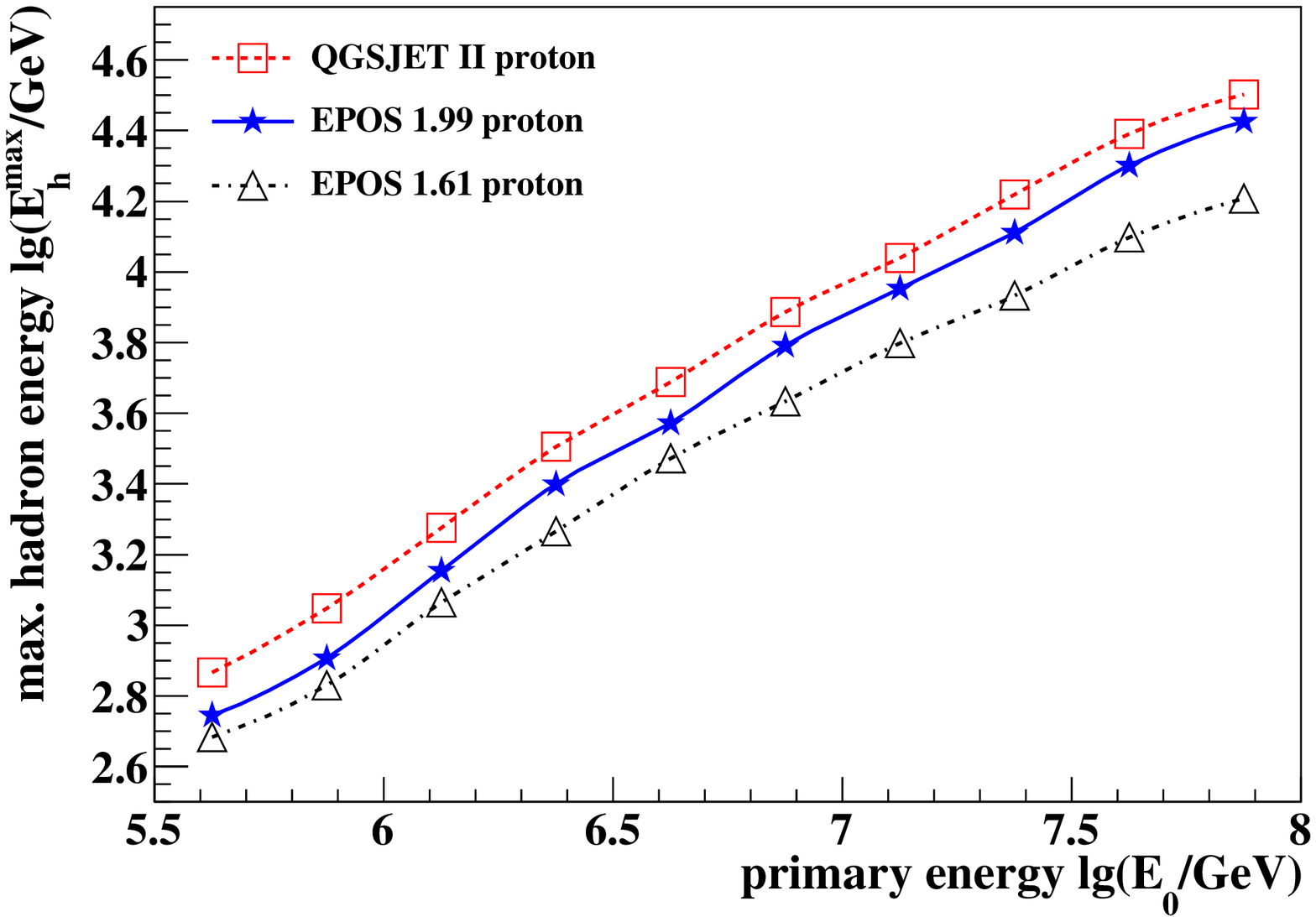} \end{center}
\vspace{11pt}
\caption{Maximum hadron energy as a function of the primary energy for proton 
induced showers using EPOS~1.99 (full line), EPOS~1.61 (dashed-dotted line) and 
QGSJET~II (dashed line) as high energy hadronic interaction models.\label{fig-Emax}}
\end{minipage}
\hfill
\begin{minipage}{0.47\textwidth}
\begin{center}
\includegraphics[width=1.\textwidth]{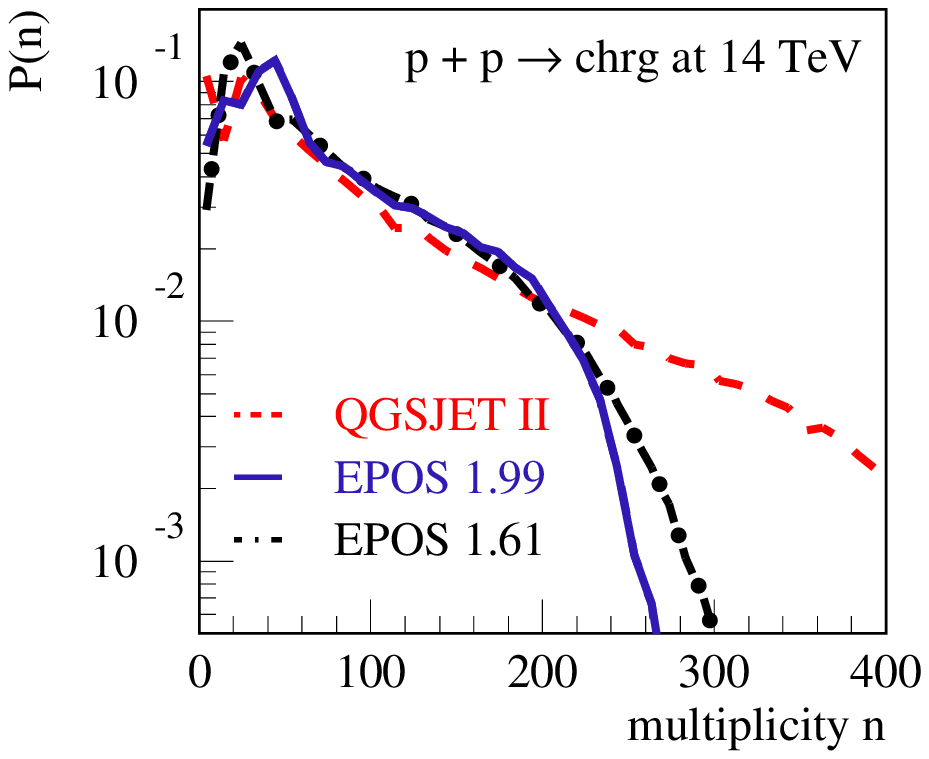}
\end{center}
\caption{Multiplicity distributions of
charged particles for inelastic events (no cut in pseudorapidity) for proton-proton collisions
at LHC for EPOS~1.99 (full), QGSJET~II (dashed), and EPOS~1.61 (dashed-dotted) hadronic interaction
models.\label{fig-LHC}}
\end{minipage}
\end{figure}

As shown on fig.~\ref{fig-zeus}, the deficit of leading
protons in EPOS~1.61 was very strong around $x_{\rm L}=0.75$. It has been
corrected in EPOS~1.99. As a consequence, EPOS~1.99 
has a reduced excitation probability at high energy compared to EPOS~1.61, 
increasing the number of protons in the forward direction and reducing 
the inelasticity. Used in air shower simulations,
the effect of the reduced cross section and inelasticity of the new EPOS version
 is clearly visible on
the maximum energy of hadrons at ground as shown fig.~\ref{fig-Emax}. The shower
being younger at ground with EPOS~1.99, the maximum energy is up to 60\% higher than 
in the previous release 1.61. The results are now close to QGSJET~II results but with
 a different slope due to a different elongation rate.

Together with a reduced number of muons at ground due to the reduced remnant 
break-up~\cite{Drescher:2007hc}, 
the results of EPOS~1.99 should be in a much better agreement with KASCADE data. 
Analysis is currently done by the KASCADE-Grande collaboration.

\subsection{Multiplicity}

The air shower data indicated that EPOS~1.61 had a too large cross section and inelasticity.
It has been corrected by an improved treatment of the non-linear effects and of the remnants. 
As a consequence, not only the results for cosmic ray experiments have been changed, but we
obtain new predictions for the LHC experiments. For instance, the multiplicity distributions of
charged particles for inelastic events (no cut in pseudorapidity) for proton-proton collision
at LHC as plotted on fig.~\ref{fig-LHC} show that EPOS~1.99 (full) has a smaller maximum multiplicity
than EPOS~1.61 (dashed-dotted) but with a larger probability for events with a small multiplicity.
QGSJET~II (dashed), which has a much larger number of parton ladder, predicts much larger fluctuations.

\section{Summary}

EPOS is an interaction model constructed on a solid theoretical
basis. It has been tested very carefully against all existing hadronic
data, also those usually not considered important
for cosmic rays. In EAS simulations, EPOS provides more muons
than other models, which was found to be linked to an increased diquark
production in both string ends and string fragmentation. To solve the problems
pointed out by the comparison with KASCADE data, the treatment of screening
effects in nuclear collisions has been improved in EPOS. The new EPOS~1.99 has 
now a reduced cross section and inelasticity compared to the previous 
EPOS~1.61 which leads to deeper shower development in better agreement with 
air shower experiments. As a consequence, we could notice a none-negligible change
for the LHC predictions. EPOS is a unique tool to test particle physics against both
accelerator experiments and air shower experiments.

\end{document}